\documentclass[lettersize,journal]{IEEEtran}

\ifCLASSINFOpdf
\else
\fi

\usepackage{times,amsmath,epsfig}
\usepackage{epstopdf}
\usepackage{color}
\usepackage{graphicx}  
\usepackage{cases}
\usepackage{algorithm}
\usepackage{algorithmic}  
\usepackage{textcomp} 
\usepackage{tikz}
\usepackage{amsthm}
\usepackage{arydshln}

\usepackage[utf8]{inputenc}
\usepackage[english]{babel}

\newtheorem{definition}{Definition}
\newtheorem{theorem}{Theorem}

\theoremstyle{plain}
  
   \ifCLASSOPTIONcompsoc 
\usepackage[caption=false,font=normalsize,labelfon
  t=sf,textfont=sf]{subfig}
  \else
\usepackage[caption=false,font=footnotesize]{subfi
  g}
  \fi

\makeatletter
  \newcommand*{\rom}[1]{\expandafter\@slowromancap\romannumeral #1@}
\makeatother

\usepackage{enumerate}
\usepackage{cite}
\usepackage{amsmath,amssymb,amsfonts}
\usepackage{algorithmic}
\usepackage{graphicx}
\usepackage{textcomp}
\usepackage{color}
\usepackage[normalem]{ulem}
\useunder{\uline}{\ul}{}
\def\BibTeX{{\rm B\kern-.05em{\sc i\kern-.025em b}\kern-.08em
    T\kern-.1667em\lower.7ex\hbox{E}\kern-.125emX}}

\newcommand{\blue}[1] {\textcolor[rgb]{0.0,0.0,1.0}{{#1}}}  

\def\0{{\mathbf 0}}
\def\1{{\mathbf 1}}

\def\p{{\mathbf p}}
\def\q{{\mathbf q}}

\def\u{{\mathbf u}}
\def\v{{\mathbf v}}

\def\x{{\mathbf x}}

\def\A{{\mathbf A}}

\def\D{{\mathbf D}}

\def\I{{\mathbf I}}

\def\L{{\mathbf L}}
\def\M{{\mathbf M}}

\def\P{{\mathbf P}}
\def\Q{{\mathbf Q}}

\def\W{{\mathbf W}}

\def\ie{{\textit{i.e.}}}
\def\eg{{\textit{e.g.}}}

\def\cE{{\mathcal E}}

\def\cG{{\mathcal G}}

\def\cO{{\mathcal O}}

\def\cT{{\mathcal T}}

\def\cV{{\mathcal V}}

\def\bdelta{{\boldsymbol \delta}}

\newcommand\norm[1]{\left\lVert#1\right\rVert}

\usepackage{multirow}
\usepackage{algorithm}
\usepackage{algorithmic}

 \ifCLASSOPTIONcompsoc 
  \usepackage[caption=false,font=normalsize,labelfon
  t=sf,textfont=sf]{subfig}
  \else
  \usepackage[caption=false,font=footnotesize]{subfi
  g}
  \fi

\begin{document}

\title{Modeling Viral Information Spreading via \\
Directed Acyclic Graph Diffusion}

\author{Chinthaka~Dinesh,~\IEEEmembership{Member,~IEEE,}
Gene~Cheung,~\IEEEmembership{Fellow,~IEEE,}
Fei~Chen,~\IEEEmembership{Member,~IEEE,}\\ 
Yuejiang~Li,~\IEEEmembership{Student Member,~IEEE,}
and~H.~Vicky~Zhao,~\IEEEmembership{Senior~Member,~IEEE}
       
\thanks{Chinthaka Dinesh and Gene Cheung are with York University, Toronto, Canada, e-mail: \{dineshc, genec\}@yorku.ca.}
\thanks{Fei Chen is with College of Computer and Data Science, Fuzhou University, Fuzhou, China, e-mail:chenfei314@fzu.edu.cn.}
\thanks{Yuejiang~Li and H.~Vicky~Zhao are with Tsinghua University, China, e-mail:lyj18@mails.tsinghua.edu.cn and vzhao@tsinghua.edu.cn.}
}



\maketitle

\begin{abstract}
Viral information like rumors or fake news is spread over a communication
network like a virus infection in a unidirectional manner: entity
$i$ conveys information to a neighbor $j$, resulting in two equally informed (infected) parties. 
Existing graph diffusion processes focus only on bidirectional
diffusion on an undirected graph. 
Instead, leveraging recent research in graph signal processing (GSP), we propose a new directed acyclic graph (DAG) diffusion process to estimate the probability $x_i(t)$ of node $i$'s infection at time $t$ given an initial infected source node $s$, where $x_i(\infty)~=~1$.
Specifically, given an undirected positive graph modeling
node-to-node communication, we first estimate its graph embedding:  a latent coordinate
for each graph node in an assumed low-dimensional manifold space via extreme eigenvectors computed using LOBPCG. 
Next, we construct a DAG
based on Euclidean distances between latent coordinates. 
Spectrally, we prove that the asymmetric DAG Laplacian matrix contains real
non-negative eigenvalues, and that the DAG diffusion converges to the all-infection vector $\x(\infty) = \1$ as $t \rightarrow \infty$. 
Simulations show that our DAG diffusion process accurately estimates the probabilities of node infection
over a variety of graph structures at different time instants. 
\end{abstract}

\begin{IEEEkeywords}
Graph diffusion, graph signal processing, graph embedding  
\end{IEEEkeywords}
%
\section{Introduction}
\label{sec:intro}

The advent of modern communications means that viral information like rumors or fake news can quickly spread within a population at an unprecedented speed~\cite{liu2014analysis}.
While an undirected graph $\cG(\cV,\cE,\W)$, containing nodes $\cV$ and edges $\cE$ with weights specified by adjacency matrix $\W$, is common to model communication between individual entity pairs---\eg, a positive weight $W_{i,j} \in \mathbb{R}^+$ of an edge $(i,j) \in \cE$ encodes the likelihood that nodes $i$ and $j$ communicate during a fixed time unit---how to model viral information spreading through $\cG$ as a process over time from an initial infected source node $s \in \cV$ to others $\cV \setminus \{s\}$ has not been studied in the \textit{graph signal processing} (GSP) literature \cite{ortega18ieee,cheung18}. 
Specifically, \textit{graph diffusion process} \cite{segarra15,thanou17, hammond2013} is traditionally studied as discretized \textit{heat diffusion}: via a \textit{bidirectional} message exchange mechanism over a graph $\cG$, the steady state has every node obtained a global average temperature. 
In contrast, viral info spreading over networks is fundamentally \textit{unidirectional}: an informed (infected) node $i$ cannot be uninfected over time, and the steady state has every node become infected\footnote{A node known to refuse spreading of viral information, and thus is never infected, can be removed from the graph \textit{a priori} in a pre-processing step.}. 
In this paper, we propose a \textit{directed acyclic graph} (DAG) based info spreading process rooted in \textit{graph embedding}~\cite{chen2022manifold,chen2022fast} to estimate the probability $x_i(t)$ of node $i$'s infection at time $t$, where $x_i(\infty) =  1$. Our DAG diffusion process differs from common diffusion models studied in social networks such as \textit{linear threshold} (LT) and \textit{independent cascade} (IS) \cite{kempe03}, which specify only how viral info may spread from a node to its neighbors, but do not estimate infection probabilities on a per-node basis as a function of time.

Specifically, interpreting an undirected positive graph\footnote{\cite{kim2012temporal} unfolded a static undirected graph into a dynamic directed graph assuming known changing network topology over time. Instead, we assume that undirected graph $\cG$ models static node-to-node communication.} $\cG(\cV,\cE,\W)$ as a set of discrete samples on a $K$-dimensional manifold \cite{xu2020manifold}---where an edge weight $W_{i,j} \propto d_{i,j}^{-1}$ conveys distance $d_{i,j} \in \mathbb{R}^+$ between nodes $i$ and $j$ on the manifold---we first estimate a \textit{latent coordinate} $\p_i \in \mathbb{R}^K$ for each node $i \in \cV$ in the manifold space. 
The node coordinates $\p_i$'s are estimated from eigenvectors that are solutions to an optimization problem minimizing one-hop neighbor distances while maximizing two-hop neighbor distances in $\cG$.  
Essentially, \textit{Euclidean distances between latent coordinates in the manifold space encapsulate network distances between nodes in graph $\cG$}.
Extreme eigenvectors of a sparse symmetric matrix can be computed in linear time using state-of-the-art numerical linear algebra algorithms, such as \textit{locally Optimal Block Preconditioned Conjugate Gradient} (LOBPCG)~\cite{knyazev2001}. 

\begin{figure}[t]
\centering
\subfloat[]{
\includegraphics[width=0.14\textwidth]{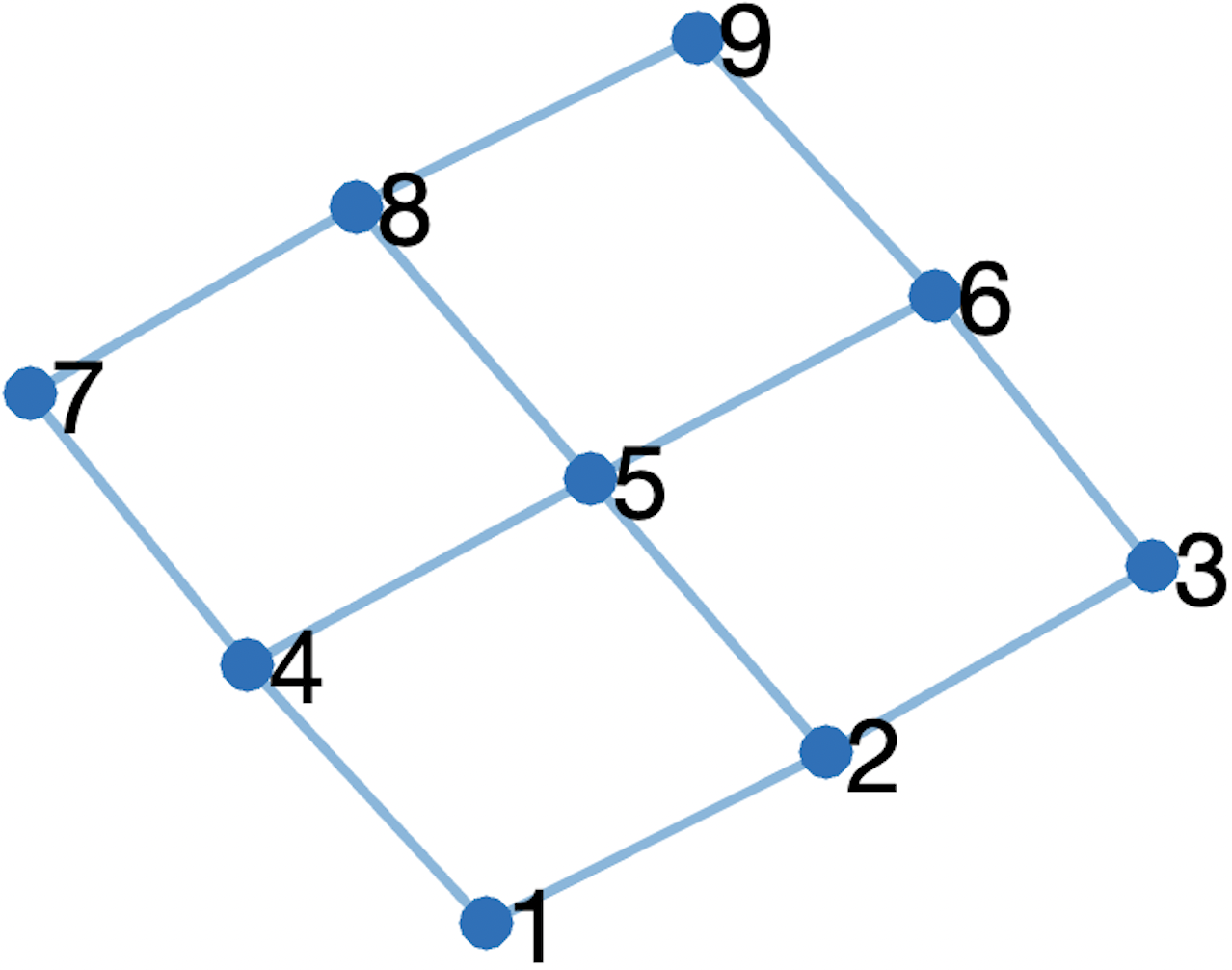}
\label{fig:pcd-on}}
\subfloat[{}]{
\includegraphics[width=0.14\textwidth]{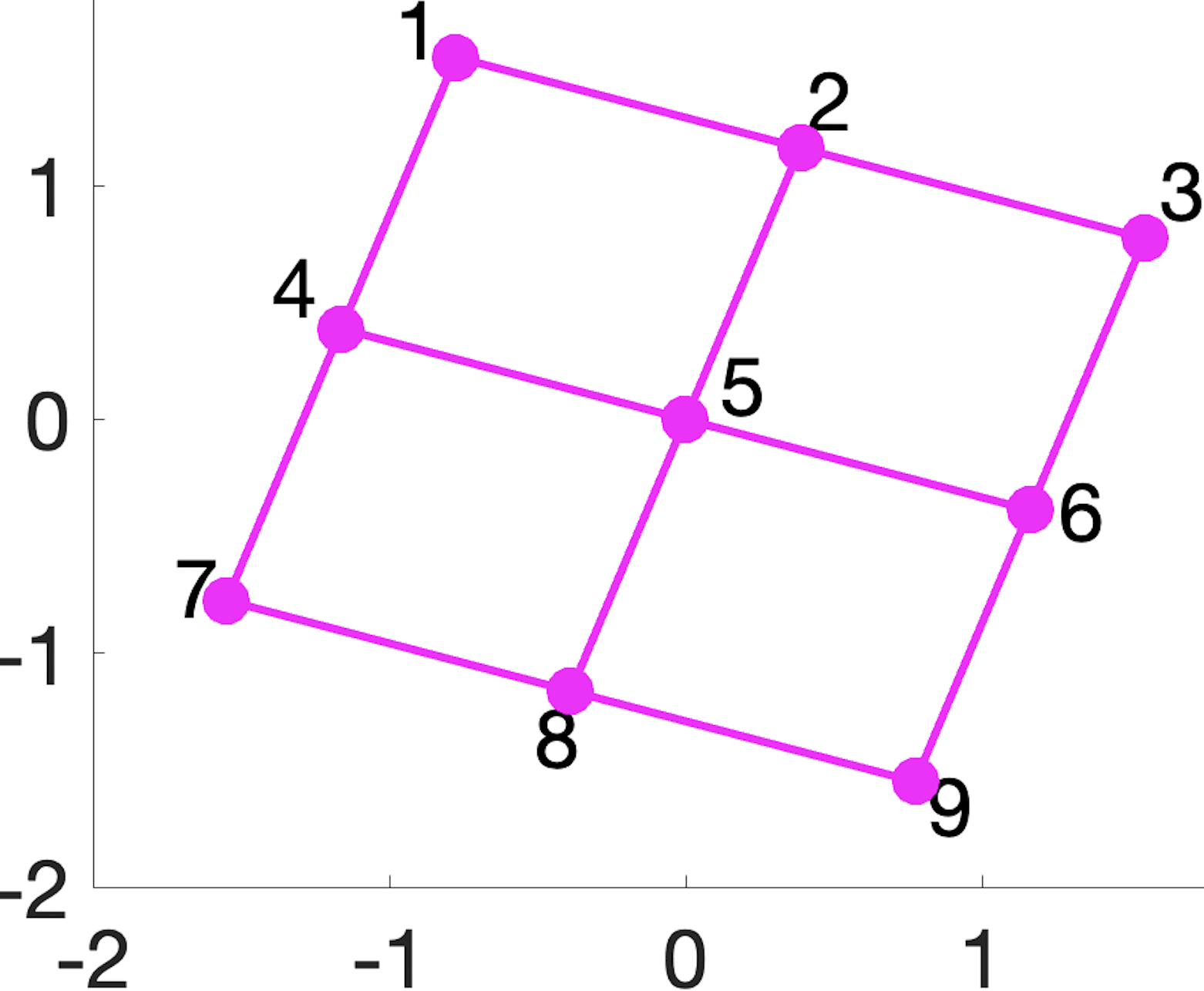}
\label{fig:pcd-off}}
\subfloat[{}]{
\includegraphics[width=0.13\textwidth]{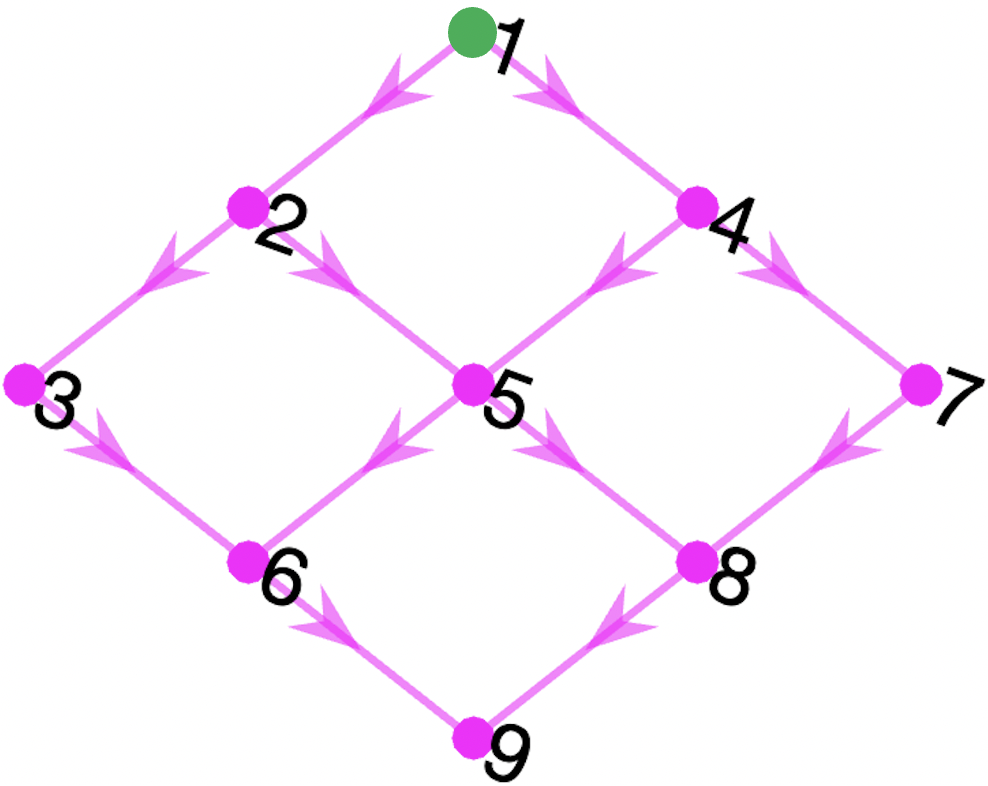}
\label{fig:pcd-off}}
\caption{Illustration of (a) a manifold graph, (b) mapping of graph nodes to 2D Euclidean coordinates, (c) a resulting DAG by assuming node 1 is the source node.} 
\label{fig:overview}
\end{figure}

Given the computed latent coordinates and an initial infected source node $s \in \cV$, we construct a \textit{directed acyclic graph} (DAG)\blue{\footnote{Recently, \cite{seifert2022causal} proposed causal Fourier analysis on DAGs, augmenting DAGs to ensure transitive closure and thus proper eigen-decomposition of the graph shift operator. In contrast, our method requires only simple eigen-analysis of the directed graph Laplacian matrix and no graph modifications.}}. 
$\bar{\cG}$ based on Euclidean distances from $s$ in the manifold space (see Fig.\;\ref{fig:overview} for an illustration). 
Spectrally, we prove that the asymmetric DAG Laplacian matrix $\L$ corresponding to $\bar{\cG}$ contains real non-negative eigenvalues, and that the DAG diffusion on $\bar{\cG}$ converges to the all-infection vector $\x(\infty) = \1$ as $t \rightarrow \infty$.  
Simulations on different graph structures show that our DAG diffusion process accurately estimates the probabilities of node infection over graphs at different time instants. 

The outline of the paper is as follows. We first present needed definitions in Section\;\ref{sec:prelim}. We discuss our DAG construction in Section\;\ref{sec:DAG} and present our DAG convergence results in Section\;\ref{sec:framework}. Finally, we present results and conclusion in Section\;\ref{sec:results} and \ref{sec:conclude}, respectively.

\section{Preliminaries}
\label{sec:prelim}
A communication graph $\cG(\cV,\cE,\W)$ has $N$ nodes $\cV$ connected by undirected edges $(i,j) \in \cE$. 
$W_{i,j} \in [0,1]$ is the non-negative edge weight encoding the ``communication strength" between node pair $(i,j)$.
\textit{Adjacency matrix} $\W \in \mathbb{R}^{N \times N}$ is real and symmetric.
Denote by $\D \in \mathbb{R}^{N \times N}$ the diagonal \textit{degree matrix}, where $D_{i,i} = \sum_j W_{i,j}$. 
The \textit{combinatorial graph Laplacian matrix} is $\L \triangleq \D - \W$. 

We define a corresponding \textit{directed acyclic graph} (DAG) $\bar{\cG}(\cV,\bar{\cE},\bar{\W})$ with the same $N$ nodes $\cV$ as $\cG$. 
$[i,j] \in \bar{\cE}$ denotes a directed edge from $i$ to $j$. 
$\bar{\W} \in \mathbb{R}^{N \times N}$ is an \textit{asymmetric} adjacency matrix, where $\bar{W}_{i,j} > 0$ is the weight of directed edge $[i,j] \in \bar{\cE}$ if it exists, and $0$ otherwise. 
We assume that $\bar{\cG}$ has a single \textit{source node} $s \in \cV$, where $\nexists \, [i,s] \in \bar{\cE}$, and there exists a path from $s$ to all other nodes $\cV \setminus \{s\}$ in $\bar{\cG}$. 
Define $\bar{\D}$ a diagonal \textit{in-degree matrix}, where $\bar{D}_{i,i} = \sum_{j} \bar{W}_{j,i}$ is the weight sum of edges entering node $i$. 
Finally, we define \textit{directed graph Laplacian matrix} $\bar{\L} \triangleq \bar{\D} - \bar{\W}^\top$.

\section{DAG Construction}
\label{sec:DAG}
\subsection{Manifold Graph}

We assume that an input undirected positive graph $\cG(\cV,\cE,\W)$ is a \textit{manifold graph}: a finite graph composed of uniformly sampled points on smooth continuous manifolds~\cite{costa2004manifold, chen2022manifold,chen2022fast}---where an edge weight $W_{i,j}$ conveys distance $d_{i,j} \in \mathbb{R}^+$ between nodes (points) $i$ and $j$ on the manifold. 
This is essentially the common \textit{manifold hypothesis} \cite{carey2017graph}: although points in a dataset are defined in a high-dimensional space, they intrinsically reside in a lower-dimensional manifold space. 
See~\cite{wright2022high} for more details on low-dimensional models for high-dimensional data. 
Mathematically, we assume that a manifold graph satisfies the following properties:
\begin{definition}
Manifold graph $\cG(\cV, \cE, \W)$ satisfies two properties:
\begin{enumerate}
\item If $W_{i,j} > W_{i,k}$, then the Euclidean distance between point pair $(i,j)$ is smaller than $(i,k)$, \ie,  $d_{i,j} < d_{i,k}$.
\item Given points $i,j,k \in \cV$ and $(j,k) \in \cE$, if $j$ and $k$ are $h$- and $(h+1)$-hop neighbors of $i$ in $\cG$, for some positive integer $h$, then $d_{i,j} < d_{i,k}$. 
\end{enumerate}
\label{def:manifold}
\end{definition}
Property 1 means $W_{i,j}$ is inversely proportional to $d_{i,j}$, \ie, $W_{i,j} \propto d_{i,j}^{-1}$.
Property 2 implies that a $2$-hop neighbor $k$ from $i$ cannot be closer to $i$ than $1$-hop neighbor $j$ if $(j,k) \in \cE$.
These properties are true if $\cG$ is a k-nearest-neighbor (kNN) graph composed of uniformly sampled points on a smooth manifold. 
We leverage these properties to compute a graph embedding.


Given an input manifold graph $\cG(\cV,\cE,\W)$, we first describe how latent coordinates $\p_i$ can be computed for each node $i \in \cV$ in Section\;\ref{subsec:coord}.
Given coordinates $\{\p_i\}_{i=1}^N$ and $\cG$, we then describe how a DAG $\bar{\cG}$ is constructed in Section\;\ref{subsec:DAG}.

\subsection{Computing Latent Coordinates}
\label{subsec:coord}

We first define matrix $\P \in \mathbb{R}^{N \times K}$, where the $i$-th row of $\P$ contains the $K$-dimensional latent vector $\p_i \in \mathbb{R}^K$ for node $i \in \cV$.
For convenience of notations, we also define the $k$-th column of $\P$ as $\q_k$---the $k$-th coordinate of the $N$ nodes. 
In order to minimize the manifold space distances between $1$-hop neighbors $(i,j) \in \cE$ connected in graph $\cG$, we first minimize the \textit{graph Laplacian regularizer} (GLR) \cite{pang17}, \ie, 
\begin{equation}
\small
\begin{split}
\min_{\P|\P^{\top}\P=\I}\text{tr} \left( \P^{\top} \L \P \right) 
&= \sum_{k=1}^K \q_k^{\top} \L \q_k \\
&= \sum_{k=1}^K \sum_{(i,j) \in \cE} W_{i,j} (q_{k,i} - q_{k,j})^2
\end{split}
\label{eq:obj1} 
\end{equation}
where $q_{k,i}$ is the $k$-th latent coordinate of node $i$, and $\I$ is an identity matrix of appropriate dimension.  
When minimizing coordinate difference in all $K$ dimensions of 1-hop neighbors $i$ and $j$ where $(i,j) \in \cE$, \eqref{eq:obj1} promotes smaller distance $d_{i,j}$ for larger edge weight $W_{i,j}$, given aforementioned property 1 of manifold graphs. 
We impose the orthogonality condition $\P^{\top}\P=\I$---\ie,  $\q_{i}^{\top}\q_{j}=\delta_{i-j}$---to ensure that the $i$-th and $j$-th coordinates are sufficiently different and thus not duplicates of each other. 
Otherwise, the dimension of the manifold space can be reduced from $K$.

Even though minimizing \eqref{eq:obj1} would minimize the weighted squared Euclidean distance $\|\p_i - \p_j\|^2_2$ between connected node pair $(i,j)$ in the manifold space, this only preserves the \textit{first-order proximity} in the graph, which may not be sufficiently representative of the global graph structure~\cite{xu2021understanding}. 
Hence, we regularize \eqref{eq:obj1} using property 2 of manifold graph.

Specifically, we define matrix $\Q$ to maximize distances between disconnected 2-hop neighbors.
Denote by $\cT_i$ the \textit{two-hop neighbor} node set from node $i$; \ie, node $j \in \cT_i$ is reachable in two hops from $i$, but $(i,j) \not\in \cE$.
The aggregate distance between each node $i$ and its 2-hop neighbors in $\cT_i$ is
$\sum_{i \in \cV} \sum_{j \in \cT_i} \|\p_i - \p_j \|^2_2$.

We can write this aggregate distance in matrix form.
For each $\cT_i$, we first define matrix $\boldsymbol{\Theta}_i \in \mathbb{R}^{N \times N}$ with entries

\vspace{-0.1in}
\begin{small}
\begin{align}
\Theta_{i,m,n} = \left\{ \begin{array}{ll}
\frac{1}{T_i} & \mbox{if}~ m = n = i ~~~\mbox{or}~~~ m = n \in \cT_i \\
-\frac{1}{T_i} & \mbox{if}~ m = i, n \in \cT_i ~~~\mbox{or}~~~ m \in \cT_i, n = i \\
0 & \mbox{o.w.}
\end{array} \right. 
\label{eq:Q}
\end{align}
\end{small}\noindent
where $T_i \triangleq |\cT_i|$ denotes the number of disconnected 2-hop neighbors.
We then define $\Q \triangleq \sum_{i \in \cV} \boldsymbol{\Theta}_i$. 
The latent coordinate optimization becomes
\begin{align}
\small
\min_{\P \,|\, \P^\top \P = \I} &~ \text{tr}(\P^{\top} \L \P) - \mu \, \text{tr}(\P^{\top} \Q \P) + \epsilon \I 
\nonumber \\ 
&= \text{tr} (\P^{\top} \underbrace{\left( \L - \mu \Q + \epsilon \I \right)}_{\A} \P ) .
\label{eq:obj2}
\end{align}
We observe that optimizing \eqref{eq:obj2} is equivalent to computing the first $K$ eigenvectors of the matrix $\A$. 
We use a state-of-the-art numerical linear algebra algorithm, LOBPCG~\cite{knyazev2001}, for this task, running in $\cO(N)$ for sparse symmetric matrices, assuming $K \ll N$. 
Parameters $\mu$ and $\epsilon$ are obtained as follows.    

As a quadratic minimization problem \eqref{eq:obj2}, it is desirable for matrix $\A=\L - \mu \Q + \epsilon \I$ to be \textit{positive semi-definite} (PSD), so that the objective is lower-bounded, \ie,  $\q^{\top} \A \q \geq 0, \forall \q \in \mathbb{R}^N$. 
We propose to first set $\epsilon = \lambda^{(2)}_{\min}(\Q)$ as the \textit{second} smallest eigenvalue of $\Q$---the Fiedler number (note that $\lambda^{(1)}_{\min}(\Q) = 0$).  
The Fiedler number---also called the \textit{algebraic connectivity} and is greater than $0$ iff the graph is connected---is a connectivity measure of a graph. 
Hence, larger $\lambda^{(2)}_{\min}(\Q)$ means that there are more 2-hop neighbors, and thus a larger $\epsilon \I$ is needed to keep $\A$ PSD.

Next, we compute $\mu > 0$ so that $\A$ is ensured to be PSD. 
By the well-known \textit{Gershgorin Circle Theorem}\footnote{Given a real square matrix $\M$, corresponding to each row $i$ is a \textit{Gershgorin disc} $i$ with center $M_{i,i}$ and radius $\sum_{j \neq i} |M_{i,j}|$. GCT states that eigenvalues of $\M$ reside inside the union of all discs. See~\cite{varga04} for details.} (GCT)~\cite{varga04}, matrix $\A$ is PSD if left-ends of all \textit{Gershgorin discs} corresponding to rows of $\A$ (disc center $c_i \triangleq A_{i,i}$ minus radius $r_i \triangleq \sum_{j \neq i} |A_{i,j}|$ for all rows $i$) are at least 0. 
Thus, we compute $\mu$ to satisfy this condition for $\A$, \ie,
\begin{align}
\small
L_{i,i} - \mu Q_{i,i} + \epsilon - \sum_{j | j \neq i} \left| L_{i,j} - \mu Q_{i,j} \right| & \geq 0, ~~\forall i .
\label{eq:discLE1}
\end{align}

Note that $L_{i,j} = -W_{i,j} \leq 0$, and $Q_{i,j} \leq 0$.
Note further that node $j$ cannot both be a $1$-hop neighbor to $i$ and a disconnected $2$-hop neighbor, and hence either $L_{i,j}~=~0$ or $Q_{i,j}~=~0$. 
Thus, we can remove the absolute value operator as
\begin{align}
\small
L_{i,i} - \mu Q_{i,i} + \epsilon - \sum_{j | j \neq i} \left( -L_{i,j} - \mu Q_{i,j} \right) &\geq 0 .
\label{eq:discLE2}
\end{align}
We set the equation to equality and solve for $\mu_i$ for row $i$, \ie, 

\begin{small}
\begin{align}
\mu_i = \frac{L_{i,i} + \sum_{j|j\neq i} L_{i,j} + \epsilon}{Q_{ii} - \sum_{j|j\neq i} Q_{i,j}} 
= \frac{\epsilon}{Q_{ii} - \sum_{j|j\neq i} Q_{i,j}} ,
\label{eq:mu}
\end{align}
\end{small}\noindent
where $L_{i,i} = - \sum_{j\neq i} L_{i,j}$.
Finally, we use the smallest non-negative $\mu = \min_{i} \mu_i$ for \eqref{eq:obj2} to ensure all Gershgorin disc left-ends of matrix $\A$ are at least $0$, as required in \eqref{eq:discLE1}.

Fig.\;2 shows selected visual results of coordinate computation for a $15$-node triangle mesh graph with $30$ unweighted edges. 
As shown, classical graph embedding methods \textit{local linear embedding} (LLE)~\cite{roweis2000nonlinear} and \textit{Laplacian eigenmaps} (LE)~\cite{belkin2001laplacian} have computed latent coordinates leading to shape distortions at the boundaries. In contrast, our eigenvector-based method (abbreviated as EV) achieved roughly equal node spacings by minimizing 1-hop neighbor distances while maximizing distances between 2-hop neighbors.

\begin{figure}[t]
\centering
\includegraphics[width=0.42\textwidth]{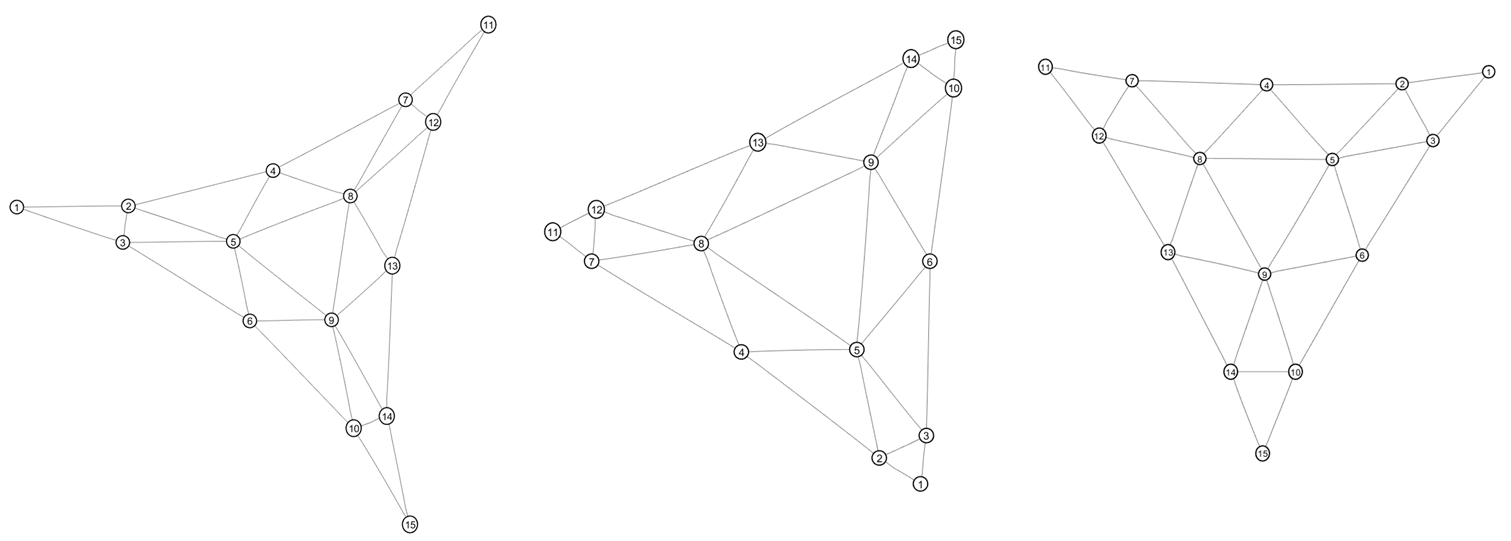}
\caption{Triangle mesh. Left: LE, middle: LLE, right: EV.} 
\label{fig:G}
\vspace{-12pt}
\end{figure}

\subsection{Constructing a DAG}
\label{subsec:DAG}

Using the computed latent coordinates $\P$ for the given undirected graph $\cG$, we define a \textit{directed acyclic graph} (DAG) $\bar{\cG}(\cV,\bar{\cE},\bar{\W})$ as follows. First, we choose a source node $s\in \cV$ from the undirected graph $\cG$. Then the DAG  $\bar{\cG}$ is defined, where a directed edge $[i,j]~\in~\bar{\cE}$ from nodes $i$ to $j$ exists iff 
\begin{enumerate}
\item nodes $i$ and $j$ are connected in $\cG$, \ie, $(i,j) \in \cE$, and
\item Euclidean distances between source $s$ and nodes $i$, $j$ satisfy $\norm{\p_s-\p_{i}}_2 < \norm{\p_s-\p_{j}}_2$.  
\end{enumerate}
See Fig.~\ref{fig:overview} for an illustrative example.

\section{DAG Diffusion Process}
\label{sec:framework}
\subsection{Spectrum of Directed Graph Laplacian for a DAG}
\label{subsec:spectrum}

We first analyze the spectrum (eigenvalues and eigenvectors) of a directed graph Laplacian matrix $\bar{\L}$ for positive DAG $\bar{\cG}(\cV,\bar{\cE},\bar{\W})$ with a single source node $s \in \cV$.

\begin{theorem}
Given a directed graph Laplacian $\bar{\L} \triangleq \bar{\D} - \bar{\W}^\top$ for a positive connected DAG $\bar{\cG}(\cV,\bar{\cE},\bar{\W})$ with $N$ nodes and one lone source node $s \in \cV$,
\begin{enumerate}
\item Eigenvalues $\{\lambda_i\}_{i=1}^N$ of matrix $\bar{\L}$ are real and non-negative, in particular, $0 = \lambda_1 < \lambda_2 \leq \ldots \leq \lambda_N$. 
\item The first left and right eigenvectors corresponding to $\lambda_1 = 0$ are $\u_1 = \bdelta_s$ and $\v_1 = \1/\sqrt{N}$.
\end{enumerate}
$\bdelta_s$ is a one-hot vector that equals to $1$ at index $s$ and $0$ otherwise, and $\1$ is a constant vector of all ones.

\label{thm:spectrumL}
\end{theorem}

\begin{proof}
Given $\bar{\cG}$ is acyclic, nodes $\cV$ can be relabeled in \textit{topological sort order}, \ie, directed edge $[i,j] \in \bar{\cE}$ implies $i < j$~\cite{kahn1962topological}. 
Hence, reordering rows/columns of directed adjacency matrix $\bar{\W}$ in topological sort order, $\bar{\W}$ becomes \textit{upper-triangular}, \ie, $\bar{W}_{i,j}~=~0, \forall i > j$, and source $s$ becomes the first node. 
$\bar{\L} \triangleq \bar{\D} - \bar{\W}^\top$ is hence \textit{lower-triangular}, meaning that its eigenvalues are exactly the diagonal entries $\bar{L}_{i,i} = \bar{D} _{i,i} = \sum_j \bar{W}_{j,i}$, which are real and  non-negative for a positive graph. 
In particular, $\lambda_1 = 0$, since $\sum_j \bar{W}_{j,s} = 0$ for source node $s$. 
Each non-source node $i \in \cV \setminus \{s\}$ has at least one in-coming edge $[j,i] \in \bar{\cE}$ with weight $\bar{W}_{j,i} > 0$, and thus $\bar{D}_{i,i} > 0$.
This implies $\lambda_i > 0, \forall i \in \{2, \ldots, N\}$. 

To show $\v_1 = \1/\sqrt{N}$ is the first right eigenvector corresponding to $\lambda_1 = 0$, we write
\begin{align*}
\small
(\bar{\L} \1)_i = (\bar{\D} \1)_i - (\bar{\W}^\top \1)_i 
= \bar{D}_{i,i} - \sum_{j} \bar{W}_{j,i} = 0, ~~~~\forall i 
\end{align*}
To show $\bdelta_s$ is the first left eigenvector for $\lambda_1 = 0$, 
\begin{align*}
(\bdelta_s^\top \bar{\L})_i &= \bar{D}_{s,s} = 0, ~~~~~~~ \mbox{for}~~ i = s \\
(\bdelta_s^\top \bar{\L})_i &= -\bar{W}_{i,s} \stackrel{(a)}{=} 0, ~~~ \mbox{for}~~ i \neq s
\end{align*}
where $(a)$ is true since source $s$ has no in-coming edges.
\end{proof}

\subsection{Convergence of DAG Diffusion}

We consider the following DAG diffusion model from source $s$ to other nodes $\cV \setminus \{s\}$ in $\bar{\cG}$. While alternative diffusion models such as LT and IS \cite{kempe03} are also possible, here we derive a simple model extending from the known heat equation \cite{smola2003, thanou17, hammond2013}. Denote by $\x(t) \in \mathbb{R}^N$ a graph signal on $\bar{\cG}$ at time $t$, where $x_i(t)$ is the probability that node $i$ is infected by time $t$. 
Time differential of signal sample $x_i(t)$ at node $i$ is
\begin{align}
\frac{\partial x_i(t)}{\partial t} &= 
\gamma \sum_{j|[j,i] \in \bar{\cE}} \bar{W}_{j,i} \max(x_j(t) - x_i(t),0) ,
\label{eq:differential}
\end{align}
where $\gamma > 0$ is a parameter.
\eqref{eq:differential} states that up-stream node $j$ in DAG $\bar{\cG}$ is worsening the infection likelihood of down-stream node $i$ at a rate proportional to: i) edge weight $\bar{W}_{j,i}$, and ii) difference between the two infection probabilities $x_j - x_i$ if $x_j > x_i$ (\ie, infection is unidirectional). 
\eqref{eq:differential} is similar to previous heat equation~\cite{smola2003, thanou17, hammond2013} but for a \textit{directed} graph.
Assuming $x_j > x_i, \forall [j,i] \in \bar{\cE}$, we can rewrite \eqref{eq:differential} in matrix form:
\begin{align}
\frac{\partial \x}{\partial t} &= \gamma \left( \bar{\W}^\top \x - \bar{\D} \x \right) = - \gamma \bar{\L} \x .
\label{eq:differential2}
\end{align}

We now state the following convergence result for DAG diffusion of DAG $\bar{\cG}(\cV,\bar{\cE},\bar{\W})$ with a single source node $s$.
\begin{theorem}
Given differential equation \eqref{eq:differential2} for a positive connected DAG $\bar{\cG}(\cV,\bar{\cE},\bar{\W})$ with $N$ nodes and one lone source node $s \in \cV$ and initial condition $\x(0) = \bdelta_s$, $\x(\infty) = \1$. 
\label{thm:convergence}
\end{theorem}

\begin{proof}
General solution to differential equation \eqref{eq:differential2} is~\cite{boyce2021elementary}
\begin{align}
\x(t) &= e^{-\gamma \bar{\L} t}\x(0) = \sum_{n=1}^N \alpha_n e^{- \gamma \lambda_n t} \v_n
\end{align}
where constants $\{\alpha_n\}_{n=1}^N$ define the initial condition for $\x(0)$. 
Given $\lambda_1 = 0 < \lambda_2 \leq \ldots \leq \lambda_N$ by Theorem\;\ref{thm:spectrumL}, we conclude that $\x(\infty) = \alpha_1 \v_1$. 

To compute $\alpha_1$, we use first left eigenvector $\u_1$ and write 
\begin{align}
\u_1^\top \x(0) &= \sum_{n=1}^N \alpha_n \u_1^\top \v_n \stackrel{(a)}{=} \alpha_1 \u_1^\top \v_1 
\nonumber \\
\alpha_1 &= \frac{\u_1^\top \x(0)}{\u_1^\top \v_1} \stackrel{(b)}{=} \frac{\bdelta_s^\top \bdelta_s}{1/\sqrt{N}} = \sqrt{N} .
\end{align}
$(a)$ is due to $\u_1^\top \v_n = 0, \forall n > 1$ (See Appendix\;\ref{sec:append}). 
$(b)$ is due to $\u_1 = \bdelta_s$ and $\v_1 = \1/\sqrt{N}$ by Theorem\;\ref{thm:spectrumL}.
Thus,
\begin{align}
\x(\infty) = \alpha_1 \v_1 =  \sqrt{N} \cdot \1 / \sqrt{N} = \1 .
\end{align}
\end{proof}

\vspace{-10pt}
\noindent
\textbf{Remark}: 
The convergence result in Theorem\;\ref{thm:convergence} is different from previous undirected graph diffusion process \cite{smola2003, thanou17, hammond2013, segarra15}; every node in DAG diffusion is eventually infected as much as source $s$, while graph nodes in undirected graph diffusion converge to an average of all initial node values.

\section{Experiments}
\label{sec:results}
 
\subsection{Experiments with Synthetic Data}
\subsubsection{Experimental Setup}
We first tested our DAG model on four manifold graphs---4-connected 2D lattice graph, 8-connected 2D lattice graph, 12-connected 2D lattice graph, and 3D lattice graph. For the experiments, we used 100, 225 and 400 nodes for 2D lattice graphs, and 108, 243 and 432 nodes for the 3D lattice graph. For a given graph, we generated 100 uniform random weights between zero and one for each edge, resulting in 1200 graphs. 
In each graph, we randomly selected a source node.       
 
To generate ground-truth signals for evaluation, we simulated viral information spreading on an undirected graph $\cG$ over time as follows. 
We first defined probability $\text{Pr}(W_{i,j})$ for each edge weight $W_{i,j}$ as $\text{Pr}(W_{i,j}) = W_{i,j}$. 
Then, for each trial $\tau$ in $\Gamma$ trials, we simulated info spreading from source $s$ at time $t$, where info was spread from $i$ to $j$ at $t$ iff: i) if $j$ was a one-hop neighbor of $i$, ii) $i$ was informed (infected) and $j$ was not, and iii) a uniform random generated number $r < \text{Pr}(W_{i,j})$, where $r\in [0,1]$. 
We recorded at each time $t$ which node was infected. 
Finally, for each designated time in $t_1, \ldots, t_M$, we computed the fraction $f_i$ of trials that each node $i$ was infected. 
The collection of fractions $f_i$'s was the simulated graph signal (probabilities) at a designated time. 

We compared our proposed DAG diffusion model against three competing schemes. 
In the first competing scheme (denoted by \textit{competitor 1}), we computed $x_i(t)$ as
\begin{equation}
x_{i}(t) = 1-e^{\frac{-\alpha t}{h_i}},
\label{eq:comp1}
\end{equation}
where $\alpha>0$ is a parameter, and $h_i$ is the number of minimum hops from source $s$ to node $i$. 
In the second competing scheme (denoted by \textit{competitor 2}), we first constructed a DAG, where a directed edge from nodes $i$ to $j$ exists iff i) $i$ and $j$ are a connected pair in $\cG$, and ii) $h_i<h_j$. 
In the third competing scheme (denoted by \textit{competitor 3}), instead of the graph embedding method in Section\;\ref{subsec:coord}, we used LLE~\cite{roweis2000nonlinear} to compute latent coordinates to construct a DAG. 
For these constructed DAGs in competitors 2 and 3, we used our DAG diffusion in Section\;\ref{sec:framework}.
We empirically tuned values for parameters $\gamma$ in~(\ref{eq:differential}) and $\alpha$ in~(\ref{eq:comp1}) to minimize MSE for the proposed method and competing schemes.

\begin{figure*}[t]
\centering
\subfloat[]{
\includegraphics[width=0.36\textwidth]{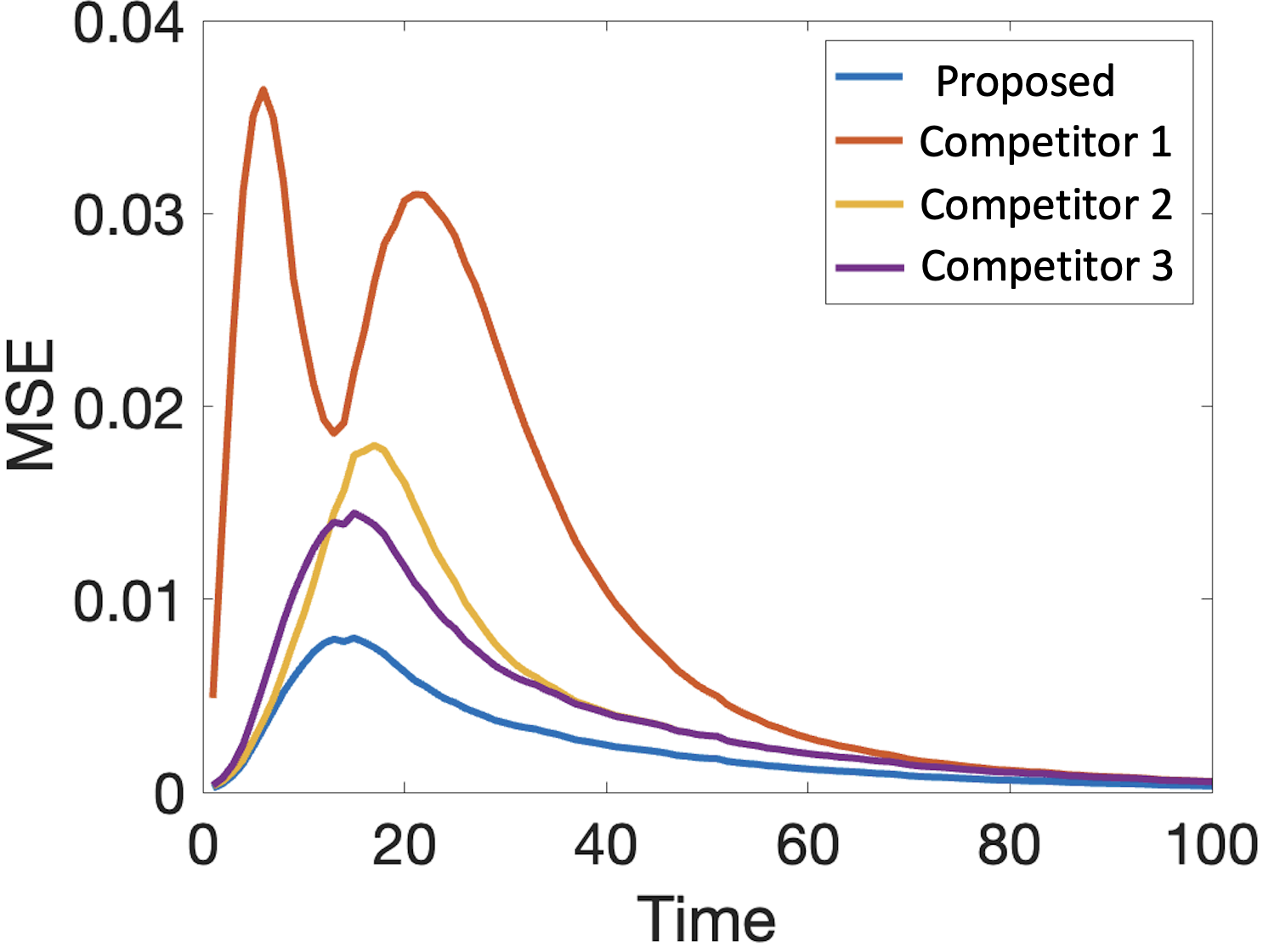}
\label{fig:pcd-on}}
\subfloat[{}]{
\includegraphics[width=0.36\textwidth]{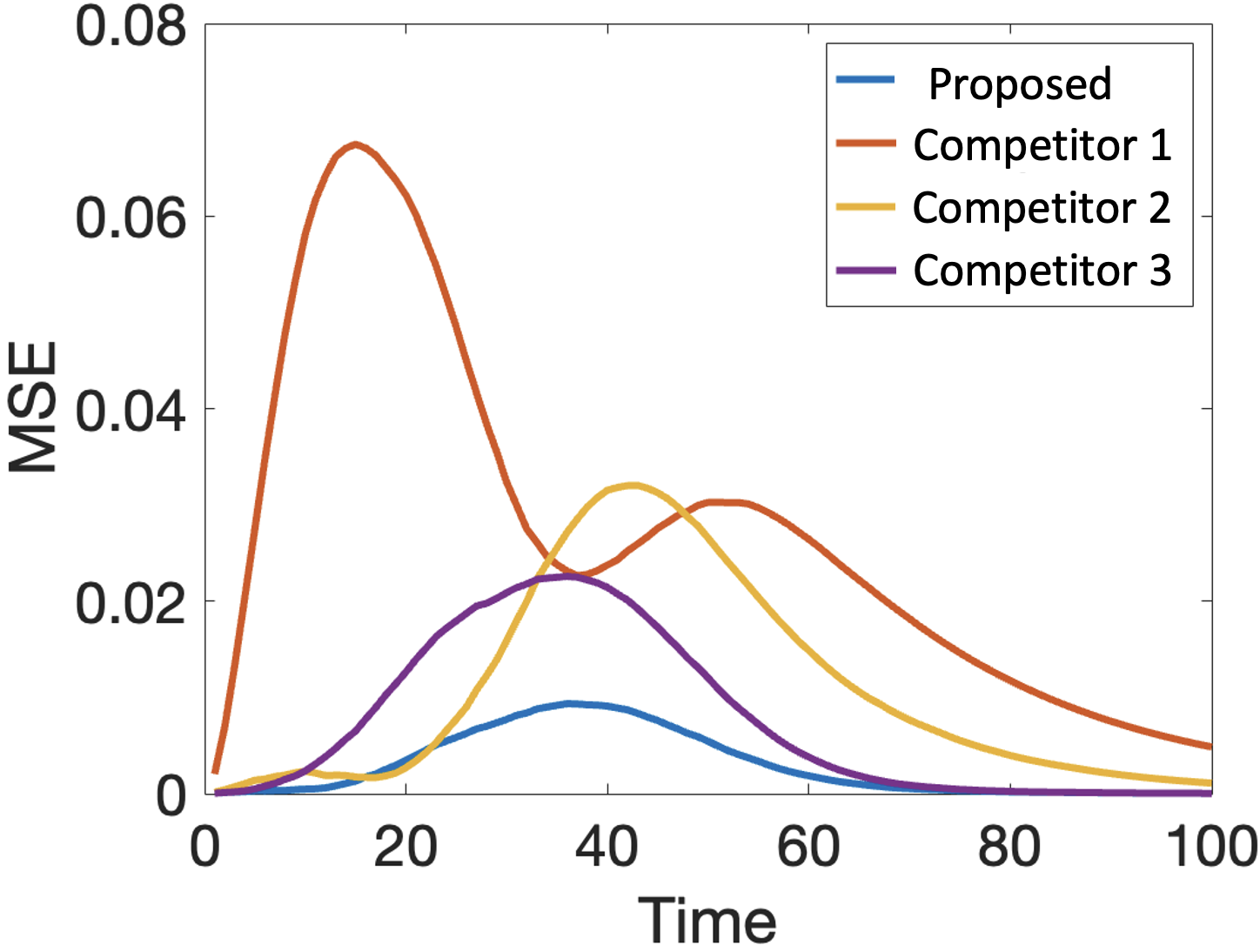}
\label{fig:pcd-off}}\\
\subfloat[{}]{
\includegraphics[width=0.36\textwidth]{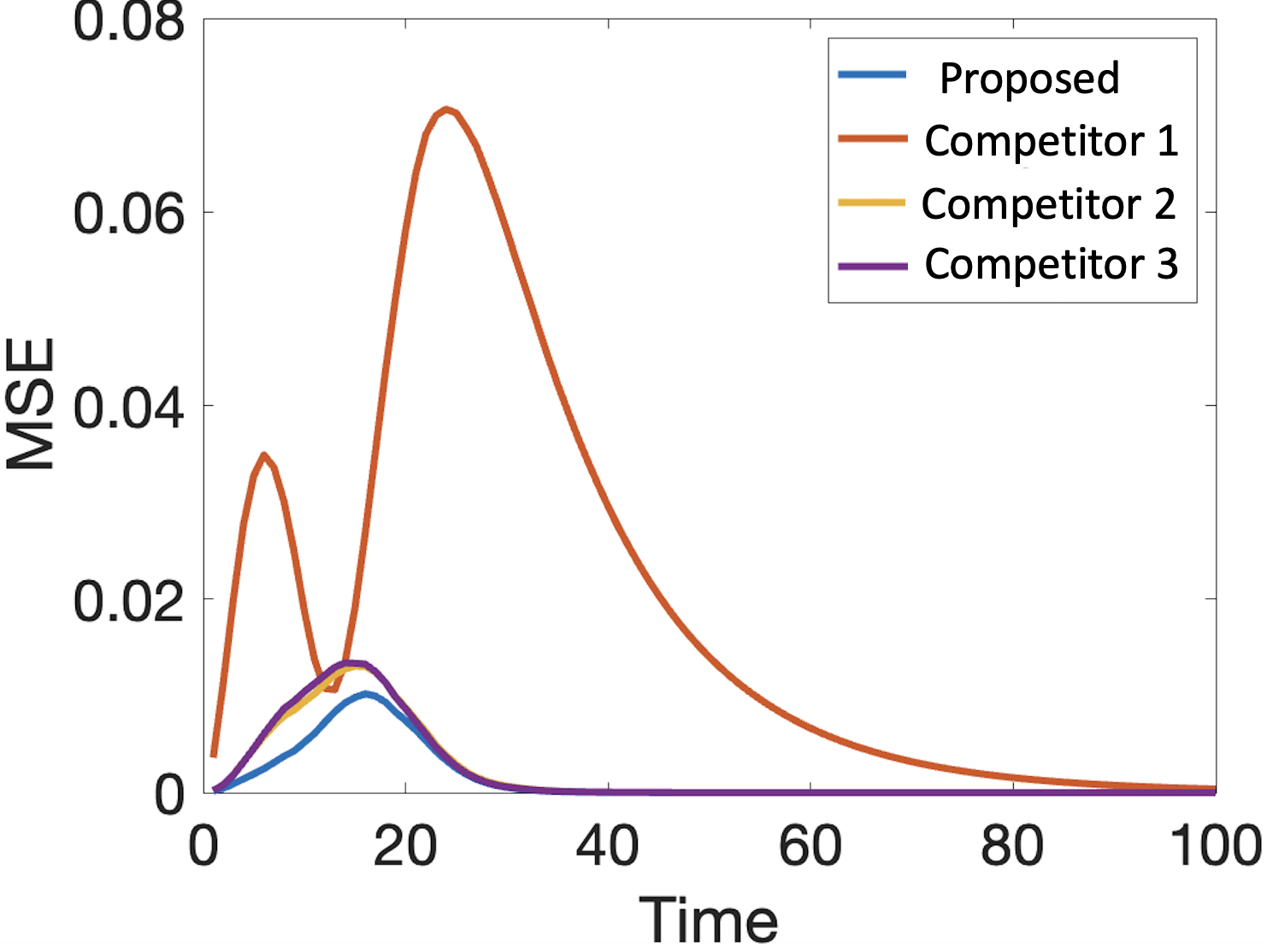}
\label{fig:pcd-off}}
\subfloat[{}]{
\includegraphics[width=0.36\textwidth]{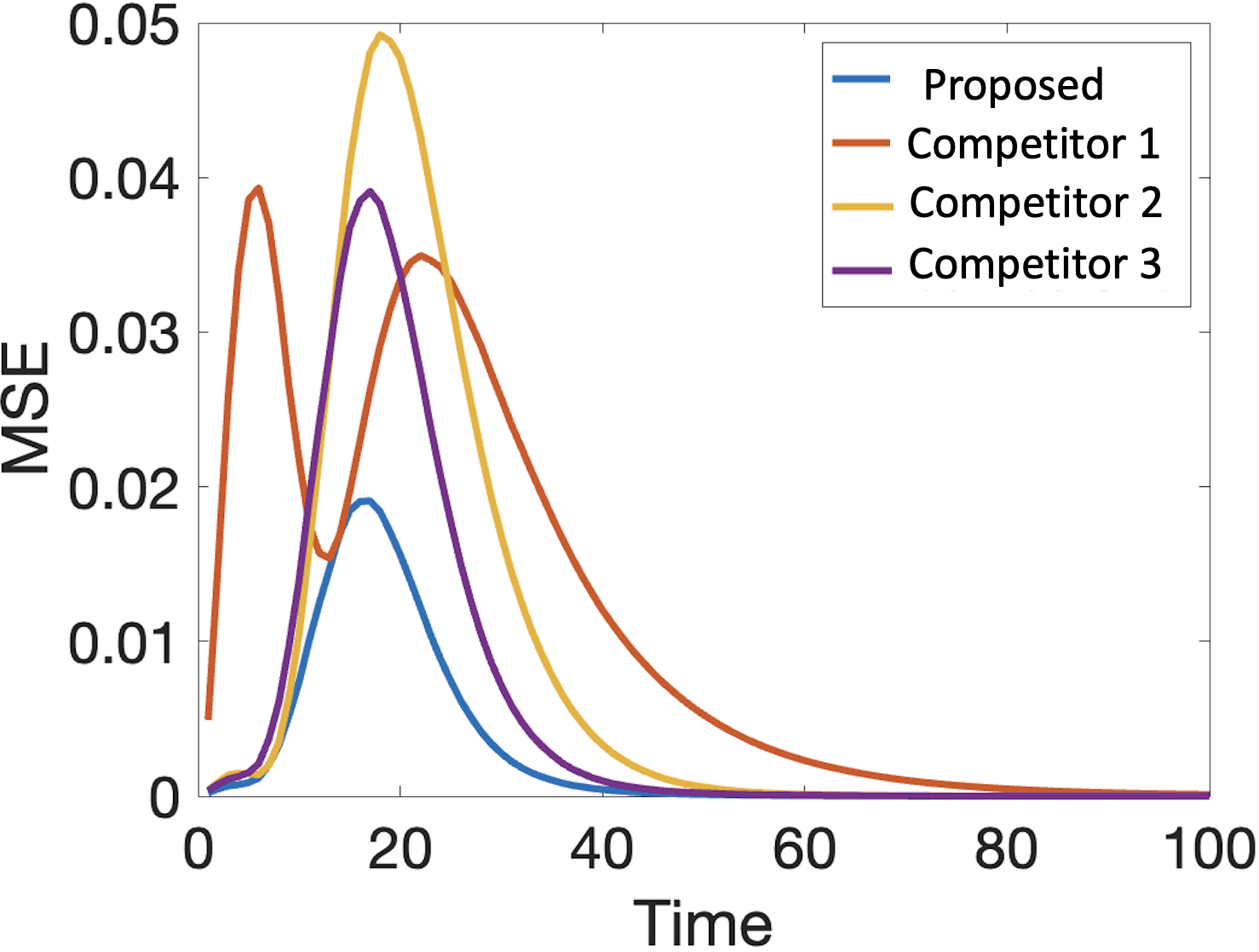}
\label{fig:pcd-off}}
\caption{Average MSE for (a) 4-connected, (b) 8-connected, (c) 12-connected 2D lattice graphs, (d) 3D lattice graphs} 
\label{fig:DAG}
\end{figure*}

\begin{figure}[t]
\centering
\subfloat[{}]{
\includegraphics[width=0.35\textwidth]{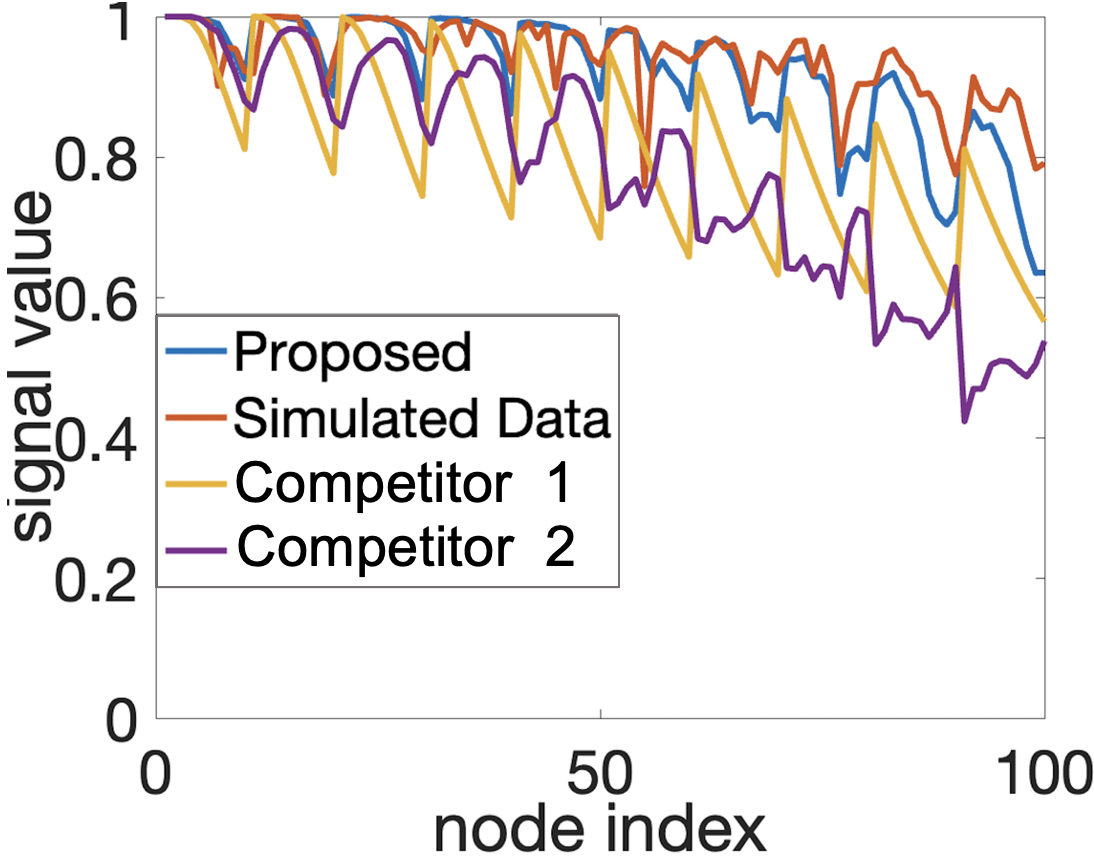}
\label{fig:pcd-off}}\\
\subfloat[{}]{
\includegraphics[width=0.38\textwidth]{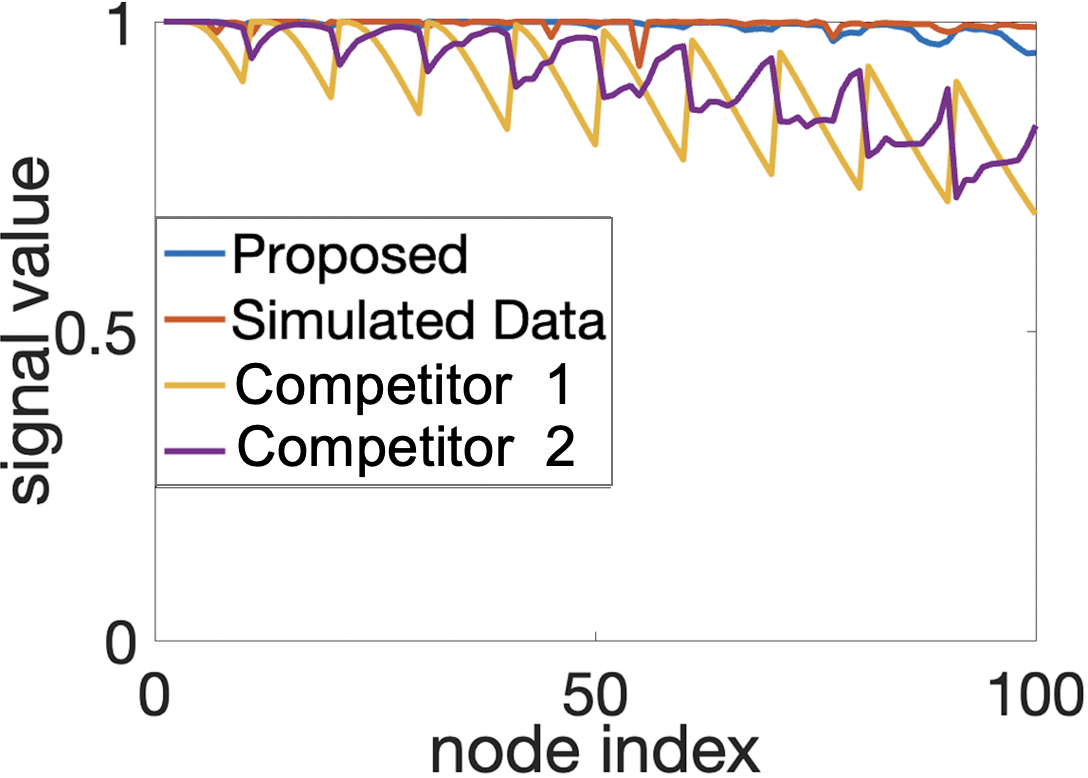}
\label{fig:pcd-off}}
\caption{Signal value $\x$ at each node of a 100-node 2D lattice graph at (a) $t=40$, (b) $t=70$. Here, we consider node 1 as the source node.} 
\label{fig:time_instants}
\end{figure}

\subsubsection{Experimental Results}

Performance of our proposed DAG diffusion and the three competitors are shown in Fig.\;\ref{fig:DAG} in terms of average MSE over time. 
For each graph, we observe that the performance of our model was noticeably better compared to the competitor 1 for each time instant on average. 
Our model was slightly better than competitors 2 and 3 for smaller time instant, and the performance gain became more pronounced for larger time instants. 
In Fig.\;\ref{fig:time_instants} we present signal value $\x$ at each node of a 100-node 2D lattice graph (4-connected) for $t=40$ and $t=70$. 
From Fig.\;\ref{fig:time_instants}, we see that our model accurately modeled info spreading over graphs at different time instants compared to competing schemes.

\subsection{Experiments with Real Data}

Though our goal is to compute probability vector $\mathbf{x}(t)$ for all nodes at time $t$, every disease infection / rumor spreading dataset describes only \textit{one instance} out of all possible graph diffusion events. 
Hence, it is impossible to validate our probabilistic diffusion model using only one real-world diffusion instance. 
Thus, we conducted a new experiment as follows. 

We used the cumulative daily Covid-19 cases\footnote{https://state.1keydata.com/} from June 1st 2020 to June 1st 2021 of seven US states. First, we constructed a fully connected undirected graph $\mathcal{G}(\mathcal{V},\mathcal{E},\mathbf{W})$ for the seven states based on Euclidean distances\footnote{https://statedistance.com} between states: 
edge weight $W_{i,j}$ between nodes $i$ and $j$ was computed as
\begin{equation}
    W_{i,j}=\exp\left(\frac{-d_{i,j}^{2}}{\sigma^{2}}\right),
\end{equation}
where $d_{i,j}$ is the distance between states $i$ and $j$, and $\sigma$ is a parameter. 
Given $\mathcal{G}$, we constructed a DAG $\bar{\mathcal{G}}(\mathcal{V},\bar{\mathcal{E}},\bar{\mathbf{W}})$ as described in Section~\ref{subsec:DAG}.

For comparison, we also computed a DAG $\bar{\mathcal{G}}^{d}$ from Covid-19 data as follows. 
We assumed the diffusion model in \eqref{eq:differential}. 
We computed the time differential of $x_{i}(t)$ as
\begin{equation}
  \frac{\partial x_i(t)}{\partial t}=x_{i}(t+1)-x_{i}(t).   
\end{equation}
Then, for each state $i$ and each time instant $t$, we obtained a set of linear equations with unknown $\bar{W}^{d}_{i,j}$. 
Since the number of equations is larger than the number of variables, we computed an optimum $\bar{W}^{d}_{j,i}$ that minimized the resulting square errors. 
Given $\{\bar{W}^{d}_{j,i}\}$, we computed the corresponding Laplacian $\bar{\mathbf{L}}^{d}$.

To evaluate the quality of our constructed DAG $\bar{\mathcal{G}}$ against DAG $\bar{\mathcal{G}}^d$ obtained from data, we compared the two corresponding Laplacian matrices $\bar{\mathbf{L}}$ and $\bar{\mathbf{L}}^{d}$ using two graph similarity metrics in the literature: \textit{relative error} (RE)~\cite{egilmez2017} and \textit{DELTACON similarity} (DCS) ~\cite{koutra2013}. 
RE between $\bar{\mathcal{G}}$ and $\bar{\mathcal{G}}^{d}$ \cite{egilmez2017} is computed as
\begin{equation}
\small
\text{RE}=\frac{\norm{\bar{\mathbf{L}}-\bar{\mathbf{L}}^{d}}_{F}}{\norm{\bar{\mathbf{L}}^{d}}_{F}},
\end{equation}
where $\norm{\cdot}_{F}$ is the Frobenius norm. DCS computes a similarity score between $\bar{\mathcal{G}}$ and $\bar{\mathcal{G}}^{d}$; it takes a value between $0$ and $1$, where value 1 means perfect similarity (see~\cite{koutra2013} for details).    

As shown in Table\;\ref{tab:graph_similarity}, $\bar{\mathcal{G}}$ computed using the proposed method is closer to $\bar{\mathcal{G}}^{d}$ than $\bar{\mathcal{G}}$ obtained from competitors 2 and 3, in terms of both metrics RE and DCS.

\begin{table}[t]
\centering
\caption{RE and DCS of $\bar{\mathcal{G}}$ obtained from competing scheme 2, competing scheme 3, and the proposed method.}
\begin{tabular}{c|c|c|c|}
\cline{2-4}
                          & Competitor 2 & Competitor 3 & Proposed \\ \hline
\multicolumn{1}{|c|}{RE}  & 0.363              & 0.372              & \textbf{0.341}    \\ \hline
\multicolumn{1}{|c|}{DCS} & 0.653              & 0.611              & \textbf{0.694}    \\ \hline
\end{tabular}
\label{tab:graph_similarity}
\end{table}

\section{Conclusion}
\label{sec:conclude}
Existing graph diffusion works focus on bidirectional diffusion on an undirected graph. We propose a new diffusion model, where we first compute a directed acyclic graph (DAG) given a source node in a communication network via graph embedding, then describe information spreading over time as DAG diffusion.
Experimental results show that our DAG diffusion model accurately estimates the probabilities of node infection at different time instants compared to competing schemes.

\appendix
\section{Appendix}
\label{sec:append}

\vspace{-0.05in}
\begin{proof}
Given left and right eigenvectors $\{\u_m\}_{m=1}^N$ and $\{\v_n\}_{n=1}^N$ of a directed graph Laplacian  $\bar{\L}$ of a positive connected DAG $\bar{\cG}(\cV,\bar{\cE},\bar{\W})$ with a single source $s \in \cV$, we prove $\u_1^\top \v_n = 0, \forall n > 1$. 
\vspace{-0.1in}
\begin{align}
\u_1^\top \bar{\L} \v_n &= \lambda_1 \u_1^\top \v_n = \u_1^\top \lambda_n \v_n \\
0 &= (\lambda_n - \lambda_1) \u_1^\top \v_n .
\end{align}
We know $\lambda_n > \lambda_1 = 0$ for $n > 1$ by Theorem\;\ref{thm:spectrumL}.
Hence, $\lambda_n - \lambda_1 \neq 0$, and therefore $\u_1^\top \v_n = 0$.
\end{proof}


\vspace{-0.1in}
\bibliographystyle{IEEEbib2}
\bibliography{ref2}

\end{document}